\documentstyle[aps,prl,floats,epsfig]{revtex}
\draft

\def\beq{\begin{equation}}
\def\eeq{\end{equation}}
\def\eea{\end{eqnarray}}
\def\bea{\begin{eqnarray}}

\def\tev{\, {\rm TeV}}
\def\gev{\, {\rm GeV}}
\newcommand{\gsim}{\lower.7ex\hbox{$\;\stackrel{\textstyle>}{\sim}\;$}}
\newcommand{\lsim}{\lower.7ex\hbox{$\;\stackrel{\textstyle<}{\sim}\;$}}

\def\slashchar#1{\setbox0=\hbox{$#1$}           
   \dimen0=\wd0                                 
   \setbox1=\hbox{/} \dimen1=\wd1               
   \ifdim\dimen0>\dimen1                        
      \rlap{\hbox to \dimen0{\hfil/\hfil}}      
      #1                                        
   \else                                        
      \rlap{\hbox to \dimen1{\hfil$#1$\hfil}}   
      /                                         
   \fi}                                         %

\begin{document}

\twocolumn[
\hsize\textwidth\columnwidth\hsize\csname @twocolumnfalse\endcsname

\title{Graviton emission from a soft brane}
\author{Hitoshi Murayama$^{a,c}$ and James D. Wells$^{b,c}$}

\address{$^{(a)}$Physics Department, University of California, 
Berkeley, CA 94720 \\
$^{(b)}$Physics Department, University of California, 
      Davis CA 95616\\
$^{(c)}$Lawrence Berkeley National Laboratory, Berkeley, CA 94720}
\maketitle

\begin{abstract}  
Theories with compact extra spatial dimensions felt only by gravity
are subject to 
direct experimental tests if the compactification volume is large.
Estimates of high-energy collider observables induced by 
graviton radiation into extra dimensions are usually given assuming
rigid branes.  This may overestimate the accessibility of these
theories.  Brane fluctuations soften the
coupling of graviton radiation, and reduce our ability to see
effects in high-energy collisions.  We calculate the size of
this suppression on single jet plus gravitons at the LHC and single
photon plus gravitons at an $e^+e^-$ linear collider. We advocate
the use of a brane softening variable as an additional parameter
when comparing theory predictions to data.

\end{abstract}  

\vspace{0.18in}

]
\setcounter{footnote}{1}


{\it Introduction:} Our universe may have $\delta$ large extra 
spatial dimensions felt by 
gravity~\cite{Arkani-Hamed:1998rs,Antoniadis:1998ig}. 
This hypothesis has given
many additional angles to attack the hierarchy problem, quantum
gravity, and the cosmological constant problem.  If the compactification
volume of the $\delta$ extra dimensions is sufficiently large, it
can have visible consequences in astrophysics, cosmology and particle
physics experiments~\cite{Arkani-Hamed:1999nn}-\cite{Kaloper:2000jb}.

High-energy
$e^+e^-$ and $pp$ collisions into photon or jet plus 
gravitons are some of the best ways that large extra dimensions can be
experimentally confirmed.  The gravitons escape the detector as a 
neutrino would, and
are inferred as missing energy in the final state.

The reduced Planck scale $M_P\equiv 1/\sqrt{8\pi G_N}$ that we usually view 
as fundamental is now a derived quantity,
\beq
M_P^2=M_D^{2+\delta}R^\delta,
\eeq
where $M_D$ is the fundamental scale of gravity, and
$R$ is the radius of toroidal compactification of the $\delta$ dimensions.
For large enough compactification radius, $M_D\simeq 1\tev$ is allowed.
The hierarchy problem, which concerns itself with why $M_P/m_W\simeq 10^{16}$
and how this ratio can be stable to quantum corrections, is now mapped
to a potentially more tractable problem of how $R$ can be stabilized
to large enough values such that all fundamental scales are roughly
equal, $M_D\simeq m_W$.

Similar to a particle in a box, the momentum of the $D=4+\delta$ dimensional
massless graviton in the $\delta$ compactified dimensions  is quantized,
\beq
p_\perp^2=\frac{\vec n\cdot \vec n}{R^2},
\eeq
where
\beq
\vec n=(n_1, n_2,\ldots, n_\delta)~~(n_i={\rm integers}).
\eeq
To an observer at low energies in the $3+1$ dimensional effective theory,
each allowed momentum in the compactified volume appears as a Kaluza-Klein
(KK) excitation of the graviton with mass $m^2=p_\perp^2$.

For any given KK graviton $G^{(n)}$, the production cross-section of 
$e^+e^-\to \gamma G^{(n)}$, for example, is
\beq
\frac{d\sigma_m}{dt}\propto \frac{1}{M_P^2}.
\eeq
Summing over all possible accessible momenta (accessible KK graviton
masses) in the extra dimensions requires integrating
\beq
\frac{d^2\sigma}{dt\, dm} = S_{\delta -1}\frac{M_P^2}{M_D^{2+\delta}}
 m^{\delta -1}\frac{d\sigma_m}{dt},
\eeq
where $S_{\delta -1}$ is the surface of a unit-radius sphere
in $\delta$ dimensions.  The $M_P^2$ factors cancel, and the
final cross-section scales as $\sigma \propto s^{\delta/2}/M_D^{2+\delta}$.
Signals can be discernible at colliders provided $M_D$ is at the
TeV scale or below.
\bigskip

{\it Brane fluctuations:} Usually calculations of final state graviton
emission such as $e^+e^-\to \gamma G^{(n)}$ and
$pp\to {\rm jet}\, G^{(n)}$ assume a rigid brane.  This implies
the coupling of a massive KK graviton is precisely the same as
a less massive KK graviton.  
A more complete formalism must ultimately
be employed to describe scattering in energy domains that probe
the flexibility of the brane.  Without such a formalism to
take into account brane fluctuations, unphysical results can arise
in calculations, such as divergent virtual contributions of 
$e^+e^-\to G^{(n)}\to \gamma\gamma$ and non-unitary 
$pp\to \gamma G^{(n)}$ production cross-sections.  Rigid cutoffs
at $M_D$ are used in many analyses to avoid these problems and
to control the size of KK graviton effects in collisions.

In this letter, we
incorporate brane fluctuations into the formalism at the beginning,
as has been suggested in refs.~\cite{Bando:1999di,Bando:2000ch}.
The simplest way to do this is to promote the extra dimensional
coordinates $\vec y=(y_1,y_2,\ldots, y_\delta)$ into fields
$\vec \phi (x)$.  The values of $\phi_i(x)$ represent the brane fluctuation
in the $y_i$ direction at the point $x$ in our $3+1$ dimensional
brane world.  The fields $\vec \phi(x)$ are Nambu-Goldstone bosons
from the spontaneous breaking of translation invariance in the $D$ 
dimensions~\cite{Sundrum:1999sj}.

Expanding to lowest order in gravity, the induced metric on the 3-brane
is
\beq
g_{\mu\nu}=\eta_{\mu\nu}+\partial_\mu \vec \phi(x)\cdot\partial_\nu\vec\phi(x).
\eeq
The kinetic terms of the $\vec \phi(x)$ fields are
\beq
 -\tau \int d^4x \sqrt{-g} \Rightarrow
  \int d^4x\left( -\tau + \frac{\tau}{2}\partial_\mu \vec\phi(x)\cdot
 \partial^\mu\vec\phi(x)\right).
\eeq
Canonically normalized fields $\vec \pi(x)$ are obtained by rescaling
$\vec\pi(x)=\sqrt{\tau}\vec\phi(x)$.

The KK reduction of the graviton field $G_{\mu\nu}$ is now
\beq
G_{\mu\nu}(x,\vec y) \propto \sum_{\vec n} G_{\mu\nu}(x)
   \exp\left({i\frac{\vec n\cdot\vec \pi(x)}{\sqrt{\tau}R}}\right).
\eeq
Substituting this into the gravity-matter interaction lagrangian,
and normal ordering the exponential, one finds a suppressed
interaction of gravity with matter,
\beq
{\cal L}\supset -\frac{e^{-\frac{1}{2}\frac{m^2}{\Delta^2}}}{M_P}
  \, G^{(n)}_{\mu\nu}\, T^{\mu\nu},
\eeq
where $m=|\vec n|/R$ is the mass of the KK graviton, 
and $\Delta^2=16\pi^2\tau/\Lambda^2$ is the
``softening scale'' derived from the brane tension $\tau$ and
the cutoff scale $\Lambda$ of a $\pi_i$-field loop.
The magnitudes and relation
of $\tau$ and $\Lambda$ are expected to be related to $M_D$, but they
are unknown.   One should be prepared for
$\Delta$ to take on almost any value.

As the KK graviton mass gets higher the coupling of the KK graviton to
matter through the energy momentum tensor $T^{\mu\nu}$ reduces.
The suppression caused by this exponential softening factor can significantly
alter the observable effects of KK graviton production at high-energy
colliders.

Insightful toy string models of large extra dimensions have been 
investigated recently~\cite{Dudas:2000gz,Cullen:2000ef,Antoniadis:2000vd}.  
The relative size of the gravitational scale to the string scale is
dictated by the strength of the Yang-Mills gauge coupling on the D-brane,
$M_D/M_S \sim 1/\sqrt{g_{\rm YM}}$.  For weakly coupled strings, the 
string scale is below the gravity scale, and string excitations 
play a significant role in the phenomenology.  The string Regge excitations
due to the finite size of 
the Standard Model (SM) particles can even overwhelm the signals of graviton
KK excitations. In this circumstance,
the cutoff that tames all high-energy observables
comes from the natural exponential suppressions inherent
in Veneziano amplitudes~\cite{Veneziano:1968yb,Hashimoto:1996kf} 
at energies near and above the string scale.

Nevertheless, brane fluctuation suppressions still can be important
in string theories.  Although
the brane fluctuations are parametrically suppressed by an exponential
factor of $g^2_{\rm YM}$~\cite{Cullen:2000ef}, the other coefficients
in the exponential are not precisely known.  
BPS brane tension calculations give $\tau_3 < M_S^4$, implying that
the exponential suppression
associated with brane fluctuations could be as significant as that from
stringy effects.  With broken supersymmetry, the tension is not presently
calculable, but if it is significantly lower than the string scale, the
brane fluctuation effects would be the most important suppression
factor in graviton emission.

Furthermore, if the gauge coupling at the string
scale is large $g^2_{\rm YM}\gsim 1$ there would not be a parametric
suppression from brane fluctuations.  In this case,
$M_D$ would be smaller than $M_S$, 
decreasing the string Regge excitations' role in providing new-physics
signatures at high-energy colliders.  This may be the 
worst-case scenario for discovering and studying low-scale
string theory,
and it maps onto our field theory description of brane
fluctuation suppressions with the identification
$\Delta^2 \simeq \tau_3/M_S^2$.  We therefore suggest that
$\Delta$ is an additional useful free parameter in evaluating
the phenomenology of large extra dimensions.

\bigskip

{\it Collider sensitivity:} Many studies have been published estimating the
sensitivity to $M_D$ and $\delta$ from virtual and external KK graviton
effects with rigid branes.  Brane fluctuation softening effects on virtual
graviton exchange have been demonstrated in 
refs.~\cite{Bando:1999di,Bando:2000ch}.  
Here we demonstrate the effects of brane fluctuations on the
results of external KK graviton production. 

\begin{figure}[bt]
\centerline{\epsfxsize=3.0truein \epsfbox{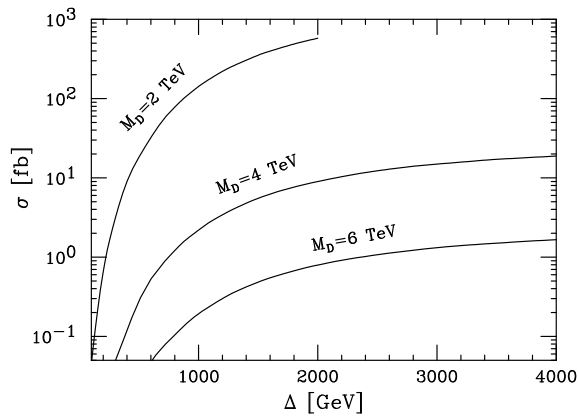}}
\vskip 0.0 cm
\caption{Signal cross-section for ${\rm jet}\, + \slashchar{E}_T$
at the LHC for $\delta=4$ extra dimensions as a function of $\Delta$, 
the exponential damping
scale for KK graviton mode couplings to matter. 
The jet is required to have
$E_{T,{\rm jet}}>1$ TeV.}
\label{soft1}
\end{figure}

In Fig.~\ref{soft1} we show the
signal cross-section for ${\rm jet}\, + \slashchar{E}_T$
at the LHC for $\delta=4$ extra dimensions as a function of $\Delta$.
The jet is required to have
$E_{T,{\rm jet}}>1$ TeV.  The three lines from top to bottom correspond
to $M_D=2$ TeV, 4 TeV, and 6 TeV.  The top line we terminate at
$\Delta=2\tev$ to be consistent with our expectation that
$\Delta\lsim M_D$. As expected the reduction of the signal is 
more than a factor of two below the rigid brane estimates ($\Delta\simeq M_D$)
for reasonable values of $\Delta\lsim M_D/2$.  For even smaller
$\Delta$ the rate drops off exponentially, making detection extremely
difficult at the LHC.

\begin{figure}[bt]
\centerline{\epsfxsize=3.0truein \epsfbox{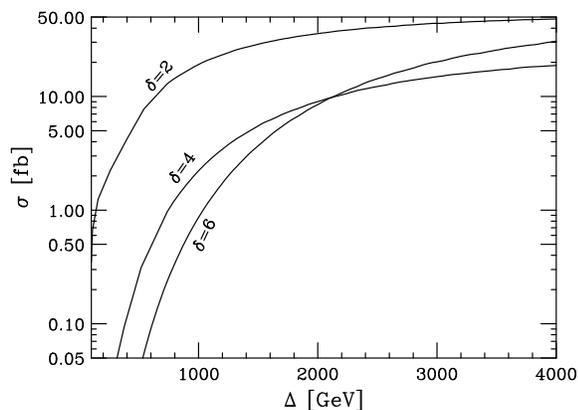}}
\vskip 0.0 cm
 \caption{Signal cross-section for ${\rm jet}\, + \slashchar{E}_T$
at the LHC for $M_D=4$ TeV as a function of $\Delta$, the exponential damping
scale for KK graviton mode couplings to matter.}
\label{soft2}
\end{figure}

In Fig.~\ref{soft2} we plot the
signal cross-section for ${\rm jet}\, + \slashchar{E}_T$
at the LHC for $M_D=4$ TeV as a function of $\Delta$.
The three lines from top to bottom correspond
to $\delta=2$, 4, and 6.  $\delta=6$ 
is initially lower than $\delta=4$ at lower
values of $\Delta$.  The ratio of the higher $\delta$ value lines to
lower $\delta$ value lines increases as $\Delta$ increases.  This is because
the density of states for higher allowed graviton masses increases
as $(\Delta R)^\delta$.  Taking into account that $R$ is a function of 
$M_D$ and $\delta$ (see eq.~1), for high-enough $\Delta$ the $\delta=6$ line
can overcome the $\delta=4$ line and cross, as it does in Fig.~\ref{soft2}.

\begin{figure}[bt]
\centerline{\epsfxsize=3.0truein \epsfbox{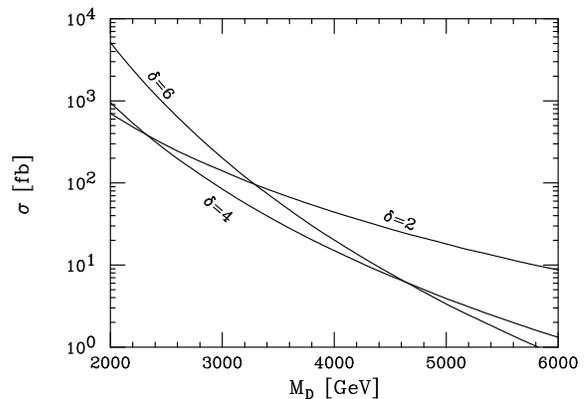}}
\vskip 0.0 cm
 \caption{Signal cross-section for ${\rm jet}\, + \slashchar{E}_T$
at the LHC for $\Delta=3$ TeV, where $\Delta$ is the exponential damping
scale for KK graviton mode couplings to matter. }
\label{soft3}
\end{figure}

In Fig.~\ref{soft3} we plot the
signal cross-section for ${\rm jet}\, + \slashchar{E}_T$
at the LHC for $\Delta=3$ TeV.
The three lines from top to bottom at low $M_D$ correspond
to $\delta=6$, 4, and 2. At higher $M_D$ the three lines reorder themselves
such that from top to bottom $\delta=2$, 4, and 6.  The reordering
of these lines is again related to the multiplicity of available KK
graviton states in the scattering process.  When $M_D< \Delta$, 
the precise value of $\Delta$ is less important and the results are close
to what are expected for the rigid brane assumption.  As $M_D$ increases
above $\Delta$ the precise value of $\Delta$ is important 
again, and the multiplicity of gravitons with unsoftened interactions becomes
$(\Delta R)^\delta \propto \Delta^\delta/M_D^{2+\delta}$.  In this regime,
the higher the number of extra dimensions $\delta$ the more suppressed
the relative number of available states and so the larger $\delta$
lines dip below the smaller $\delta$ lines.

\begin{figure}[bt]
\centerline{\epsfxsize=3.0truein \epsfbox{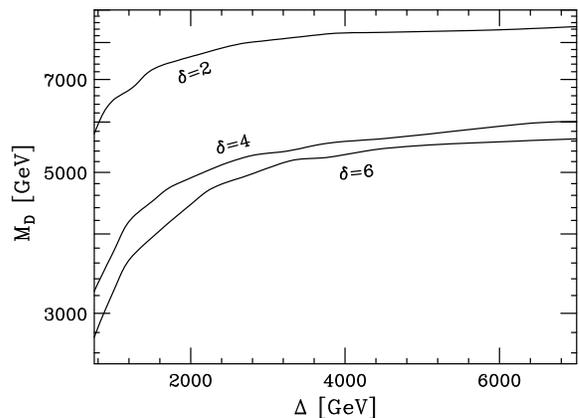}}
\vskip 0.0 cm
 \caption{$M_D$ sensitivity (as explained in text) 
at the LHC with 100 ${\rm fb}^{-1}$ as
a function of $\Delta$, the exponential damping
scale for KK graviton mode couplings to matter.  }
\label{lhcp}
\end{figure}

In Fig.~\ref{lhcp} we plot the
$M_D$ sensitivity at the LHC with 100 ${\rm fb}^{-1}$ as
a function of $\Delta$. To be clearly above background we require
at least 260 events with
$E_{T,{\rm jet}}>1\tev$~\cite{Giudice:1999ck}.  
The estimated $M_D$ sensitivity with
large $\Delta$ is similar to what is obtained for a rigid 
brane~\cite{Giudice:1999ck}.  However, with lower values of $\Delta$ the 
$M_D$ sensitivity of the LHC drops precipitously.  

Finally, we plot the
$M_D$ sensitivity at a $1\tev$ $e^+e^-$ linear collider 
assuming 500 ${\rm fb}^{-1}$ as
a function of $\Delta$.  This plot was obtained by requiring that the
photon have transverse energy above $300\gev$ and total energy below
$450\gev$ to overcome backgrounds~\cite{Giudice:1999ck}.  
We also assume 90\% polarization
of the $e^-$ beam to further reduce the $e^+e^-\to \nu\bar \nu \gamma$
background from $WW$ fusion.  Again, the sensitivity reduces dramatically
for lower values of $\Delta< M_D$.  

One advantage of the linear collider
is the precise center-of-mass energy of the $e^+e^-$ collisions.
With data at several different center-of-mass energies, one could in principle
fit the total rates to the exponential softening and determine
the $\Delta$ mass scale.  It would likely 
take careful measurements at more than
two different center-of-mass energies to resolve unambiguously the values of
$\Delta$, $M_D$ and $\delta$.  Angular distributions and photon energy
distributions at a single energy are of limited value.

\begin{figure}[bt]
\centerline{\epsfxsize=3.0truein \epsfbox{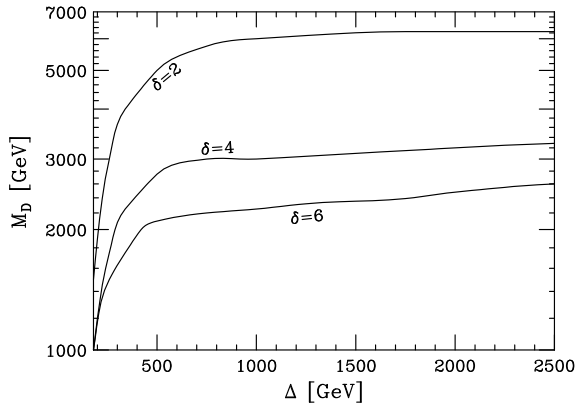}}
\vskip 0.0 cm
 \caption{$M_D$ sensitivity (as explained in text)
at a $1\tev$ $e^+e^-$ linear collider 
assuming 500 ${\rm fb}^{-1}$ as
a function of $\Delta$, the exponential damping
scale for KK graviton mode couplings to matter.}
\label{nlcp}
\end{figure}

\bigskip
{\it Conclusion:}
We have demonstrated that the ability to discover large extra dimensions
via external graviton emission is diminished if the SM 3-brane has 
low tension compared to the fundamental gravity scale.  This induces large
brane fluctuations, softening the effective coupling between the 
extra-dimensional graviton and SM states living on the brane.

The softening of graviton couplings to matter only occurs for Kaluza-Klein
modes with mass above the scale $\Delta$, or in other words significant
momentum into the extra dimensions.  Astrophysical 
constraints~\cite{Arkani-Hamed:1999nn,Cullen:1999hc,Hall:1999mk} are
derived from SM particle interactions with KK modes of
very small mass compared to reasonable expectations for $\Delta$.
Therefore, astrophysical constraints are likely not affected much
by brane fluctuations.  Nevertheless, there exist more complex
compactification
schemes which are unconstrained by astrophysical experiments,
and yet still can be probed effectively
at high-energy colliders~\cite{Kaloper:2000jb}.  
In these cases, we expect that 
a full
complement of collider and astrophysical tests will be needed to
understand the geometry of
compactification, the properties of the SM brane, and the nature of the
more fundamental theory.

\bigskip
%
\noindent
{\it Acknowledgments: }
HM was supported in part by the U.S.~Department of
Energy under Contract DE-AC03-76SF00098, and in part by the National
Science Foundation under grant PHY-95-14797. JW was
supported by the Department of Energy and the Alfred P. Sloan Foundation.

\def\Journal#1#2#3#4{{#1} {\bf #2}, #3 (#4)}
\def\add#1#2#3{{\bf #1}, #2 (#3)}

\def\NPB{{\em Nucl. Phys.} B}
\def\PLB{{\em Phys. Lett.}  B}
\def\PRL{{\em Phys. Rev. Lett.}}
\def\PRD{{\em Phys. Rev.} D}
\def\PR{{\em Phys. Rev.}}
\def\ZPC{{\em Z. Phys.} C}
\def\SJNP{{\em Sov. J. Nucl. Phys.}}
\def\AnnP{{\em Ann. Phys.}}
\def\JETPL{{\em JETP Lett.}}
\def\LMP{{\em Lett. Math. Phys.}}
\def\CMP{{\em Comm. Math. Phys.}}
\def\PTP{{\em Prog. Theor. Phys.}}

\end{document}